\documentstyle[11pt,aaspp4]{article}

\lefthead{Cardall and Fuller}
\righthead{BBN and Discordant Deuterium Determinations}

\begin{document}

\title{Big-Bang Nucleosynthesis in Light of 
	Discordant Deuterium Measurements}

\author{Christian Y. Cardall and George M. Fuller}
\affil{
Department of Physics, University of California, San Diego,
La Jolla, California 92093-0319}

\begin{abstract}
Two dimensional concordance plots involving the baryon-to-photon
ratio, $\eta$, and an effective number of light neutrinos,
$N_{\nu}$, are used to discuss
the overall consistency of standard big-bang nucleosynthesis
in light of recent determinations of the primordial
deuterium ($^2$H) abundance.
Observations of high-redshift Ly-$\alpha$ clouds have provided
discordant $^2$H/H determinations: one cloud with $^2$H/H high
compared with the previously accepted upper limit on 
($^2$H + $^3$He)/H; and one system with a
significantly lower upper bound on $^2$H/H 
than those previously obtained. 
The high value of $^2$H/H agrees well with the
current observationally-inferred
primordial abundances of $^4$He and $^7$Li for $N_{\nu}=3$.
The low value of $^2$H/H does not fit well with
the current observationally-inferred
primordial abundance of $^4$He for $N_{\nu}=3$. In addition, if the 
low value of $^2$H/H
is indicative of the primordial deuterium abundance, then
significant depletion of $^7$Li in old, hot Pop II halo stars
is probably required to obtain a concordant range of $\eta$,
for {\em any} effective number of neutrino flavors.
The use of conservative ranges for the primordial abundances of
$^2$H, $^4$He, and $^7$Li allow success of the standard picture 
for $N_{\nu}=3$.
\end{abstract}


Considerations of big bang nucleosynthesis (BBN) comprise our best
probe of the physics of the early universe. 
Primoridal nucleosynthesis calculations successfully 
predict the relative abundances
of the lightest elements over some ten orders of magnitude.
However, as claimed determinations of primordial abundances 
have grown
in precision in the recent past, a number of different 
points of view have arisen regarding the overall consistency of the
standard big-bang nucleosynthesis picture. The fact that proponents
of these different points of view sometimes 
ascribe high confidence
levels to their conclusions makes the situation 
all the more confusing. Recent deuterium measurements 
in high-redshift Ly-$\alpha$ clouds 
(\cite{tytler96}; \cite{rugers96}; 
see also \cite{songaila94}; \cite{carswell94}; 
\cite{hogan95b}) prompt a reexamination of
the problem.

Though we recognize the basic success of the theory of BBN,
we have little confidence that the commonly assigned 
uncertainties in the observationally-inferred primordial
abundances of the light elements reflect the potentially large
systematic errors in these quantities.
Sasselov \& Goldwirth (1995), for example, have emphasized
how systematic errors in the observationally-inferred 
primordial $^4$He abundance conceivably can be much larger
than the well-determined statistical errors
(\cite{olive95a}; \cite{olive95b}) in this measurement.
Arguably, the primordial $^7$Li 
abundance can also be statistically well
determined from the so-called ``Spite plateau'' 
(\cite{spite82}; \cite{molaro95}; \cite{ryan96}), but 
significant depletion and/or production of $^7$Li may 
have occurred (\cite{deliyannis95}; \cite{vauclair95};
\cite{chaboyer94}).
Any inference of the
primordial $^3$He abundance is heavily dependent on models
of stellar and galactic evolution 
(e.g., \cite{dearborn96}; \cite{hogan95a}). As for $^2$H, 
at least three significantly
different determinations of its primordial abundance exist, 
using two completely different methods.

Nevertheless, it is clearly desirable 
and possible to use BBN considerations to 
constrain $N_{\nu}$ and the baryon-to-photon ratio  $\eta$. 
 Here we follow common practice (e.g. 
\cite{walker91}) and use $N_{\nu}$ to represent all
relativistic particle degrees of freedom 
(except for photons and electrons) extant at the nucleosynthesis
epoch. In this sense, $N_{\nu}$ parametrizes the expansion rate. 
Given our lack of precise knowledge of the primordial 
abundances at present, it is useful to explore 
the leverage each elemental abundance has on both 
``concordance'' and the values of $N_{\nu}$ and $\eta$. A useful
tool for this purpose is a two-dimensional concordance 
plot, in which abundance contours are plotted in the 
$\eta$-$N_{\nu}$ plane. (Similar plots have been used in the
past in other contexts, for example by Kang \& Steigman (1992)
in a study of the effects of neutrino degeneracy on the outcome of
BBN.)
These plots give insight into both the 
degree of compatibility of determinations 
of the primoridal abundances of $^2$H, $^4$He, and $^7$Li amongst 
themselves, and the compatibility of these abundances with the
assumption of three families of light neutrinos and 
no additional relativistic degrees of freedom,
i.e. $N_{\nu}=3$.

In this work, the primordial abundances of $^4$He, $^2$H,
and $^7$Li were computed with the Kawano (1992) update of
the Wagoner (1969, 1972) code. We have used the world average
neutron lifetime of 887.0 s (Montanet et al. 1994), and the 
reaction rates of Smith, Kawano, \& Malaney (1993). 
In addition, we have employed a small correction to the $^4$He 
mass fraction (+0.0031, roughly independent of $\eta$) 
which arises from 
higher order effects in the weak rates and the use of a smaller
time step (\cite{kernan94}). We have not accounted for the
reaction rate uncertainties in the calculated BBN yields 
(\cite{krauss95}), except to consider the effect on the calculated
abundance of $^7$Li, the element for which the theoretical 
uncertainty is 
by far the largest of any of the elements produced by BBN. 
However, given the
potential systematic uncertainties in the inferred primordial 
abundances,
we do not attempt to obtain
precise numerical limits on $N_{\nu}$ or $\eta$ with associated
statistically meaningful confidence levels. 
Our goal is simply to indicate the ranges of $N_{\nu}$ and  
$\eta$ suggested by various deuterium measurements.

Consider Fig. 1, in which the range
of primordial $^2$H obtained by Hata et al. (1995b) is employed:
$1.7 \le d_5 \le 6.2$, which is assigned a 95\% C.L.
(In our discussion of the primordial light element abundances,
we use the following definitions:
$d_5 \equiv ^2$H/H $\times 10^5$;  $l_{10}
\equiv ^7$Li/H $\times 10^{10}$; and $Y$ is 
the primordial mass fraction of $^4$He.)
This range of $d_5$ is determined from solar system and 
interstellar medium measurements, and models of chemical
evolution. While this specific range was determined for 
$N_{\nu}=3$, we will take it as representative of the most stringent
inferences of the primordial deuterium abundance that can be
obtained from ``local'' measurements.
Recently adopted 
ranges for the observationally-inferred primordial
abundances of $^4$He and $^7$Li are  (Olive 1995)
$0.223 \le Y \le 0.245$ and $0.7 \le l_{10} \le
3.8$, respectively.  These ranges include $\pm 2\sigma$ statistical
and $\pm 1\sigma$ systematic errors. Included in the systematic
error in the range for $^7$Li/H is an allowance for a factor of two
depletion to get from the primordial abundance to the measured
value on the Spite plateau.

The primordial abundances of $^4$He, $^2$H,
and $^7$Li will 
constrain bands in the $\eta$-$N_{\nu}$ plane,
whose overlap yields an `allowed region.' 
For the range of primordial $^2$H/H adopted in Fig. 1, 
it is clear that the standard
model value of $N_{\nu}=3$ is only marginally compatible 
with the conventional 2$\sigma$ upper limit
of $Y=0.245$. This is one of the central contentions 
of the Hata et al. (1995a) paper: they claim a best fit of 
$N_{\nu}=2.1 \pm 0.3$,
with $N_{\nu}=3$ ruled out at 98.6\% C.L.  
This potential conflict was noted earlier by Kernan \& Krauss
(1994).
A philosophically different 
statistical approach, however, is taken by Copi et al. (1995b).
In that work, a Bayesian analysis is employed in which it 
is taken as a prior assumption that there are at least three
neutrino flavors. This procedure necessarily
yields an upper limit on $N_{\nu}$ which is greater
than three. They then argue that the $^4$He abundance has been
systematically underestimated.

Another method of measuring $^2$H involves the observation of 
Lyman-$\alpha$ systems in the line of sight to QSOs via their
absorption spectra. This method has the potential to more accurately
determine the primordial $^2$H abundance, since the QSO absorption
systems are at high redshift and have very low metallicity
(\cite{hogan95b}; \cite{malaney96}; \cite{jedamzik96}). Fig. 2
shows the range of primordial $^2$H inferred from the first such system
in which $^2$H has been claimed to be detected (\cite{hogan95b}; 
see also \cite{songaila94}; \cite{carswell94}; \cite{rugers96}).
We have used the latest range cited by Hogan (1995b):
 $1.5 \le d_4 \le 2.3$. This 
reflects only $1\sigma$ errors; 
since statistical and
systematic errors are combined in this range, 
we are not able to extend  the statistical
errors to the $2\sigma$ level. 
The adopted $^4$He and $^7$Li primordial abundance
bands in Fig. 2 are the same as in Fig. 1. 
From this figure it is clear that
$N_{\nu}=3$ fits well with the adopted primordial abundance
ranges
of all three elements plotted. Interestingly, 
this range for deuterium
is essentially that picked out if $N_{\nu}=3$ is assumed, and only
the conventional observationally-inferred ranges of primordial
$^4$He and $^7$Li abundances 
are used to determine the appropriate range of
$\eta$ (\cite{fields95}; \cite{olive95c}).

This high range of primordial $^2$H abundance is not without 
problems, however. Such
a high value of $^2$H/H may be difficult to reconcile with 
local measurements of $^3$He. Since $^2$H burns to $^3$He, 
stellar and galactic evolution would have to
destroy much more $^3$He than is sometimes thought possible to
reduce its abundance to that observed today in the solar system.
Whether enough $^3$He can be destroyed to accommodate high 
primordial $^2$H abundances is still a matter of current debate
(e.g. \cite{dearborn96}; \cite{hogan95a}).
In addition, the lower
cosmic baryon density implied by a higher $^2$H abundance aggravates
the alleged ``baryon catastrophe.'' The ``catastrophe'' is that
cluster masses derived from X-ray measurements 
seem to imply a cosmic baryon 
density about a factor of three higher than that allowed 
by BBN, for a critical density universe 
(\cite{white93}; \cite{white95}). In addition, the existence
of a significant
baryonic component of the dark halos surrounding spiral galaxies 
is suggested by recent gravitational microlensing 
studies (\cite{griest96}). This could also be a potential problem for this
high range of D/H,
because the low baryon density corresponding to high values of 
primordial deuterium leaves room for only a limited amount
of baryonic dark matter, probably not enough to account for
most of the halo mass (\cite{hogan95b}).    

Another detection of $^2$H 
in a QSO absorption system has been made by a separate
group (\cite{tytler96}). 
Compared with the high value of $^2$H/H determined 
from the Songaila et al. (1994) object,
this measurement yields a {\em low} value of $^2$H/H:
$^2$H/H $= 1.5-3.4 \times 10^{-5}$, which represents
a $2\sigma$ statistical plus $1\sigma$ systematic error
range.
As Fig. 3 shows, this range of
$^2$H  is  
inconsistent with $N_{\nu}=3$ for the range of $^4$He used
in Figs 1-2.

In addition,
this range for $^2$H appears to be 
compatible with the $^7$Li abundance only when appeal is made
to a factor of at least two or so depletion of primordial 
$^7$Li to arrive at the
Spite plateau value of $^7$Li/H. This conclusion cannot be
made with certainty, however, since the theoretical uncertainty
in $^7$Li/H can allow for a marginal agreement between the
Spite plateau value of $^7$Li/H and the new, low $^2$H/H
value. 
In Fig. 4 the Tytler et al. (1996) range of $^2$H/H
is plotted along with the upper limit on $\eta$ that
arises from
the upper limit on $^7$Li/H determined solely from
the Spite plateau, without any allowance for depletion:
$l_{10} \le 2.2$ (Olive 1995). 
In this figure no concordance is obtained, {\em for any 
effective number of neutrinos}, for the central values of 
the reaction rates entering the calculation. The figure also
shows, however, that the
uncertainties in these rates allow for a sliver of a concordance.
Nevertheless, it is clear that allowance for some $^7$Li
depletion provides much better agreement between $^7$Li/H
and the Tytler et al. (1996) measurement of $^2$H/H.
Acceptance of the range of primordial deuterium shown in
Figs. 3-4 would thus likely
require rethinking of: (1.) our 
understanding of the $^4$He and $^7$Li
primordial abundances, and/or: (2.) the 
simplifying 
 assumptions of the
standard BBN picture, including entropy homogeneity, three light
($\lesssim 1$ MeV) neutrinos,
and negligible net lepton number.
The potential conflict between the 
$^4$He abundance and 
$N_{\nu}=3$ was previously noted based on ``local'' deuterium 
measurements alone (\cite{kernan94}; \cite{hata95a}), 
and solutions reflecting the 
two options above have already found expression. 
Krauss \& Kernan (1995) emphasize the need for reconsideration
of systematic uncertainties, and
Copi et al. (1995b) suggest that the $^4$He abundance has
been systematically underestimated. On the other hand, Hata et al.
(1995a) hint at non-standard early universe neutrino physics.
 
Non-standard neutrino physics can ``fix'' the primordial 
$^4$He abundance rather easily in the case of the Hata et al. (1995b)
range of the deuterium abundance. However, 
non-standard neutrino physics is less likely to be a solution
to the conflict between the $^2$H/H and $^7$Li/H 
primordial abundances that arises if the Tytler, Fan, \& Burles
(1996) range
of deuterium is adopted. This is because altering the neutrino 
physics affects the production of $^4$He much more strongly than
the production of the other light elements formed in the big bang.
For example, consider the possibility that the cosmic neutrino seas
contain net lepton number
(i.e., the neutrinos have a nonzero chemical potential $\mu$). 
Calculations with the Kawano (1992) code reveal that an
electron neutrino
degeneracy parameter ${\mu}_e / kT \sim 0.03$ 
is sufficient to bring $^4$He, $^2$H, and $^7$Li into good
agreement for $N_{\nu}=3$ {\em if} $d_5\sim 5$. On the other hand, for
$d_5 \lesssim 2.5$, an electron neutrino degeneracy parameter
${\mu}_e / kT \sim 0.4$ is required to bring $^2$H and $^7$Li 
into concordance for $N_{\nu}=3$. However, $\mu / kT \sim 0.4$
yields $Y \approx 0.13$, which could only (maybe) be brought back up
to the observed $^4$He abundance with extreme fine tuning involving
large values of the other neutrino degeneracy parameters 
(\cite{kang92}).
  
Therefore, the acceptance of the low Tytler, Fan, \& Burles (1996) 
value of $^2$H/H would strongly suggest 
that significant ($\gtrsim$ factor of 2) 
depletion of $^7$Li may be required
to obtain concordance between the observationally-inferred
primordial abundances of $^2$H, $^4$He, and $^7$Li. 

Two recent observations of $^6$Li in old, 
hot  Pop II halo stars (\cite{smith93}; \cite{hobbs94}) 
have been cited as possible evidence against $^7$Li depletion in
this class of objects (\cite{steigman93}; \cite{copi95a}). 
This is because $^6$Li is destroyed 
at considerably lower temperature than is $^7$Li.
Therefore, if rotation-induced turbulent mixing is invoked to
effect significant $^7$Li depletion (\cite{pins92}; \cite{chaboyer94}),
we would expect that $^6$Li would suffer even greater, if not
complete, depletion. If an appeal is made to turbulent mixing 
depletion of $^7$Li in old halo stars, it would have to be argued
that $^6$Li is produced by galactic cosmic rays or {\em in situ}
by stellar flares (\cite{smith93}; \cite{deliyannis95b}).
On the other hand, Vauclair \& Charbonnel (1995) have claimed 
that another possible mechanism of 
$^7$Li depletion---mass loss via stellar winds---would not 
destroy $^6$Li, and would thus be a ``safe'' method of 
$^7$Li depletion. 
It is perhaps suggestive that the upper bound
to the range of $^7$Li/H 
obtained by Vauclair \& Charbonnel (1995) 
with this mechanism is $4.0 \times 10^{-10}$, 
in reasonable agreement with what Fig. 3 would predict, given the new 
deuterium determinations.

Allowing the possibility of significant depletion of $^7$Li brings
to mind models of inhomogeneous BBN, in which $\eta$ is 
allowed to have spatial variations on scales either smaller or
larger than the horizon scale at the epoch of BBN 
(\cite{alcock87}; \cite{applegate88}; \cite{mathews90};
\cite{thomas94}; \cite{jedamzik94}; \cite{jedamzik95};
\cite{copi95c}; \cite{gnedin95}). Higher $^7$Li abundance yields
at a given $\eta$ compared with standard BBN are a usual feature
of inhomogeneous schemes (but see Jedamzik et al. (1994)).
However, a feature of almost all inhomogeneous 
models is a high average $^2$H/H yield relative 
to a homogeneous model.
Inhomogeneous models tuned to give low $^2$H/H yields
usually overproduce $^4$He relative to a homogeneous case.
An inference of a universally low primordial $^2$H/H (in the range of 
Tytler et al. (1996)), which we have shown would
imply a requirement for significant $^7$Li depletion, would probably
be incompatible with inhomogeneity,
or would at least significantly narrow the allowed parameter
range for inhomogeneous models.
 Should the Tytler et al. (1996)
value of $^2$H/H turn out to be close to the primordial value
(e.g., only such a low $^2$H/H is detected along many lines of
sight and among many Ly-$\alpha$ clouds), probably
we could conclude that the {\em assumption} of homogeneity
is a good one, even though there is no face value concordance
in $\eta$ and $N_{\nu}$ for the usually adopted $^4$He abundance
and the Spite plateau $^7$Li abundance. 

Of course, if {\em both}
the Rugers and Hogan (1996) and Tytler et al. (1996) values
of $^2$H/H are instrinsically primordial---the difference not
being a result of differential chemical evolution or observational
selection effects---then inhomogeneity in $\eta$ at the epoch of
BBN is established. Indeed, just such intrinsic variations would
be expected in inhomogeneous isocurvature models 
(\cite{jedamzik95}).

The basic success of big-bang nucleosynthesis is remarkable.
However, at present there are discordant determinations of the
primordial deuterium  abundance.
As claimed inferences of 
primordial abundances have become more
refined, some discomforts have surfaced. A two-dimensional concordance
plot in the $\eta$-$N_{\nu}$ plane is a 
good tool to study these issues,
as it clearly displays both the overall consistency of the theory
and the leverage each element exerts on 
constraining $N_{\nu}$ and $\eta$.
The high value of $^2$H/H $\approx 2 \times 10^{-4}$ 
(\cite{rugers96})
measured in one QSO absorption system fits
well with the currently popular observationally-inferred
determinations of the $^4$He and
$^7$Li abundances for $N_{\nu}=3$. However, the low baryon 
density implied by this high range of $^2$H/H may conflict with
determinations of the baryon content of x-ray clusters and 
the massive halos surrounding spiral galaxies.
Should the low value $^2$H/H $= 1.5-3.4 \times 10^{-5}$ 
(\cite{tytler96}) obtained from another QSO
absorption system persist in future observations, it could
require rethinking of: (1.) our 
understanding of the $^4$He and $^7$Li
primordial abundances, and/or: (2.) the 
simplifying assumptions of the
standard BBN picture, which include homogeneity, three light
($\lesssim 1$ MeV) neutrinos,
and negligible net lepton number. 
In particular, it is unlikely that the 
incompatibility of the ``Spite plateau'' value of $^7$Li/H 
with $^2$H/H $\approx 1.5-3.4 \times 10^{-5}$
can be resolved in a believable way with non-standard
early universe neutrino physics or inhomogeneous BBN.
Instead, it would probably be necessary to invoke significant 
($\gtrsim$ factor of two)
depletion of $^7$Li in old, hot Pop II halo stars in order to obtain a
concordant range of $\eta$, for {\em any} value of $N_{\nu}$. 

The use of conservative ranges for the inferred primordial 
abundances of the light elements produced in the big bang 
confirms the basic success of the theory for $N_{\nu}=3$, as 
is shown in Fig. 5. The abundance of primordial deuterium could 
be anywhere between the high and low values suggested by measurements
of QSO absorption systems. 
Consideration of systematic effects may suggest that 
the upper limit on the $^4$He abundance
could be as high as $Y=0.255$ (\cite{sasselov95}). If rotation-induced
mixing has occurred in old, hot Pop II halo stars, 
primordial $^7$Li may have been depleted by a 
factor of ten (\cite{pins92}; 
\cite{chaboyer94}; \cite{deliyannis95}). Contours corresponding to 
the high value of primordial $^7$Li/H implied by such severe 
depletion do not even appear in the range of parameter space
included in Fig. 5. 
Classic predictions of the theory remain essentially intact.
For example, the effective number of
light species present at nucleosynthesis is most likely less than four
(especially if $^2$H/H turns out to be low),
and the existence of both baryonic and non-baryonic dark matter 
is suggested. Even with quite 
conservative
ranges of the primordial abundances, big-bang nucleosynthesis 
retains predictive power.

{\em Note Added.} Another measurement of $^2$H/H 
in a QSO absorption
system has been made recently by Burles and Tytler (1996). They
determine $^2$H/H in this object to be 
$1.5 \le d_5 \le 4.2$, where this
range includes $\pm 2\sigma$ statistical error $\pm$ systematic 
error. The range of $^2$H/H implied by combining this measurement
with the previous Tytler et al. (1996) measurement is 
$1.7 \le d_5 \le 3.5$ (Burles and Tytler 1996); 
therefore our conclusions regarding the impact of `low' $^2$H/H
remain unchanged. 
Burles and Tytler (1996) argue that these two
measurements taken together, along with observational difficulties
in existing measurements of QSO absorption systems that 
could be interpreted as yielding
`high' $^2H$/H (\cite{rugers96}; \cite{carswell96}; \cite{wampler96}), 
indicate a low primordial deuterium abundance.

\acknowledgments

We are indebted to S. Burles and D. Tytler for explanation of and 
numerical results from their recent measurements of D/H
in Ly-$\alpha$ systems. 
We thank K. Jedamzik and L. Krauss for useful
conversations. This work was supported by 
NSF Grant PHY-9503384 and a NASA Theory Grant at UCSD.

\begin{figure}
\plotone{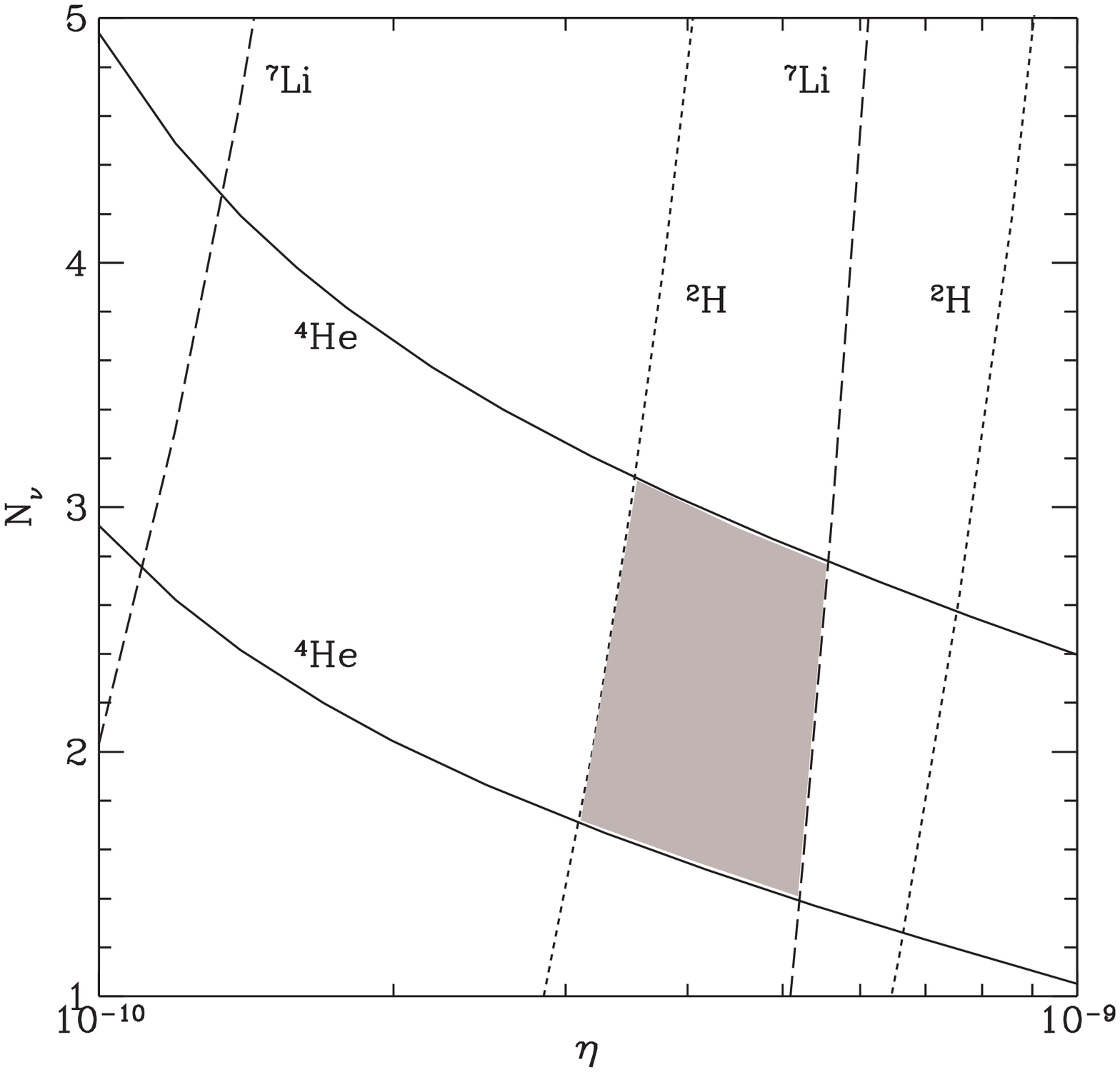}
\caption{The allowed region in the 
	$\eta-N_{\nu}$ plane for
	the range of $^2$H (dotted lines) as determined  
	by Hata et al. (1995b)
	from solar system and
	interstellar medium measurements:
	$1.7 \le d_5 \le 6.2$. This range is quoted as
	a 95\% C. L. interval. For $^4$He (solid lines) and 
	$^7$Li (dashed lines) we have 
	used 2$\sigma$ ranges on the quoted statistical error, 
	plus or minus the quoted systematic error: 
	$0.223 \le Y \le 0.245$, $0.7 \le l_{10} \le
	3.8$ (Olive 1995). Both $^7$Li contours correspond to
	the upper bound; a sufficiently large lower bound would
	bifurcate the $^7$Li band, but this does not occur
	for the present lower bound. This range of $^7$Li allows
	for up to a factor of two depletion.}
\end{figure}

\begin{figure}
\plotone{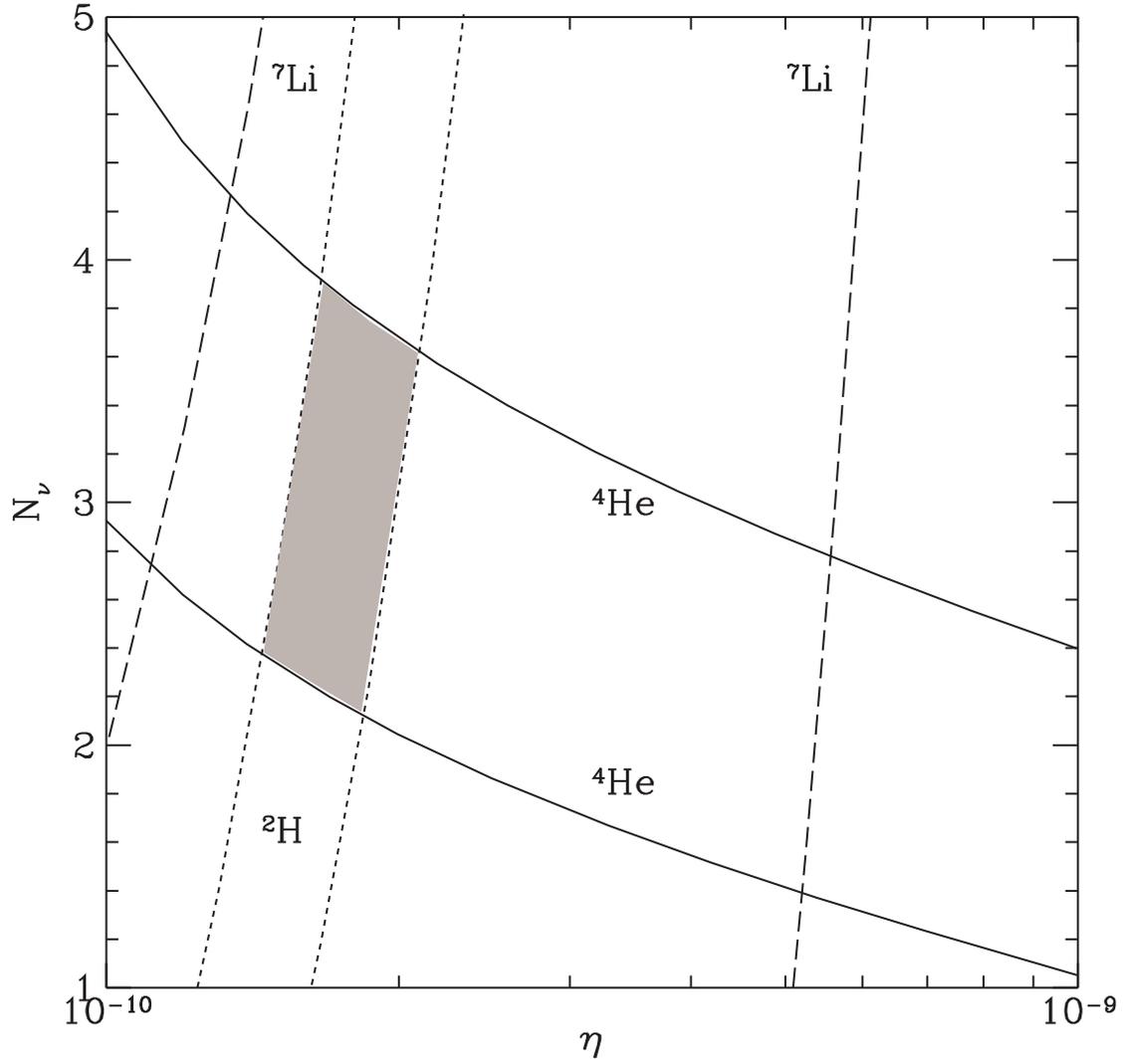}
\caption{The allowed region in the 
	$\eta-N_{\nu}$ plane for
	the range of $^2$H (dotted lines) as determined 
	from a QSO absorption system: $1.5 \le d_4
	\le 2.3$ (Rugers \& Hogan 1996). This is a 1$\sigma$ range.
	The ranges for $^4$He (solid lines) and $^7$Li (dashed
	lines) are as in Fig. 1.}
\end{figure}

\begin{figure}
\plotone{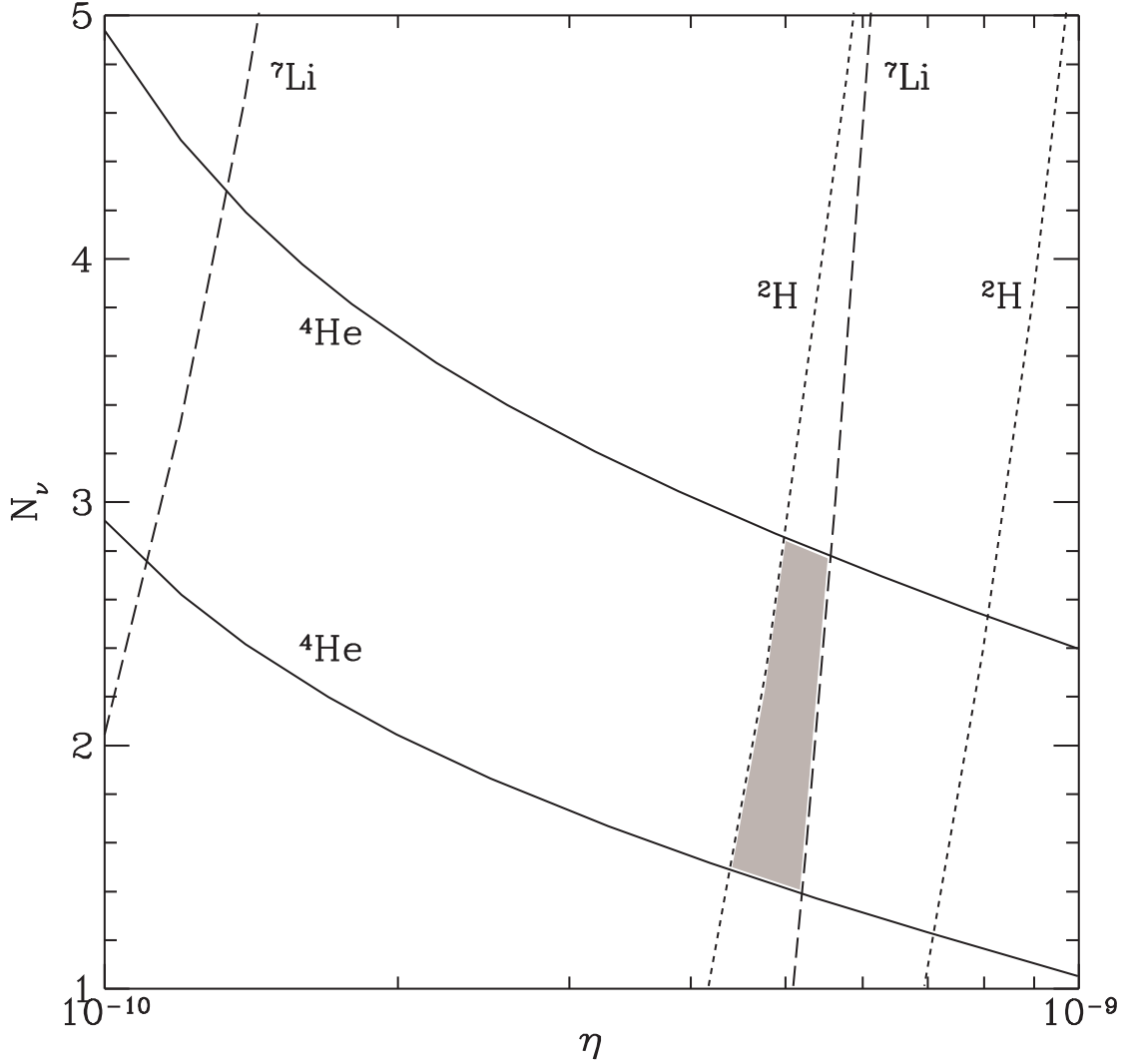}
\caption{The allowed region in the 
	$\eta-N_{\nu}$ plane for
	the range of $^2$H (dotted lines) as determined 
	from a QSO absorption system: 
	$1.5 \le d_5 \le 3.4$ (Tytler, Fan, \& Burles 1996).
	This reflects 2$\sigma$ ranges on the quoted statistical 
	error, plus or minus the quoted systematic error.
	The ranges for $^4$He (solid lines) and $^7$Li (dashed
	lines) are as in Fig. 1.}
\end{figure}

\begin{figure}
\plotone{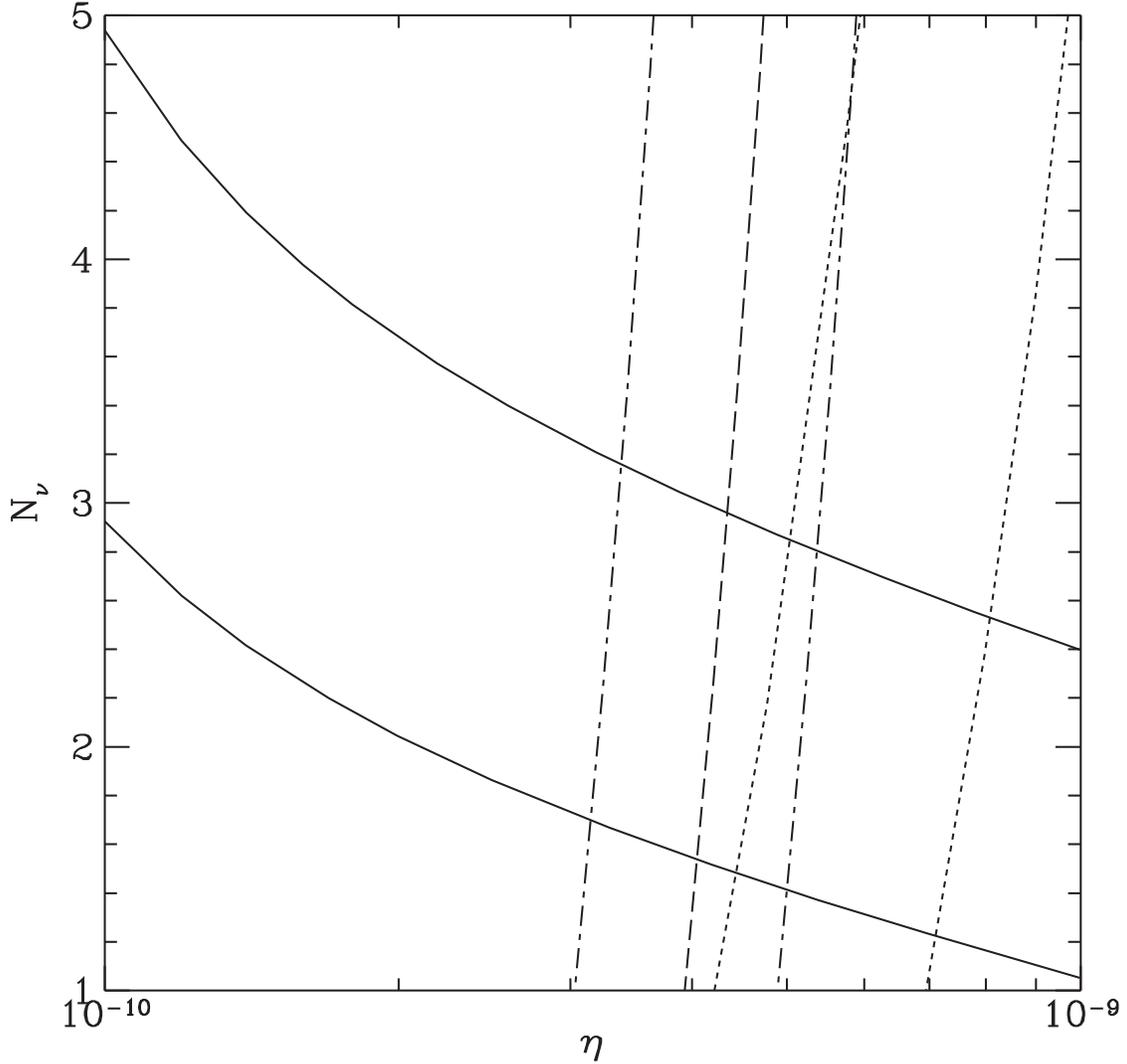}
\caption{The allowed region in the 
	$\eta-N_{\nu}$ plane for
	$^2$H (dotted lines) as in Fig. 3, $^4$He 
	(solid lines) as in
	Figs. 1-3, and the ``Spite plateau'' value of
	$^7$Li/H, with no allowance for depletion.
 	Only the $^7$Li contour yielding the upper bound on
	$\eta$ is shown: $l_{10} \le
	2.2$ (Olive 1995). The dashed line indicates
	the $l_{10}=2.2$ contour that arises from use of the central
	values of the reaction rates, and the dot-dashed
	lines indicate the range allowed by the uncertainties
	in these rates.}
\end{figure}

\begin{figure}
\plotone{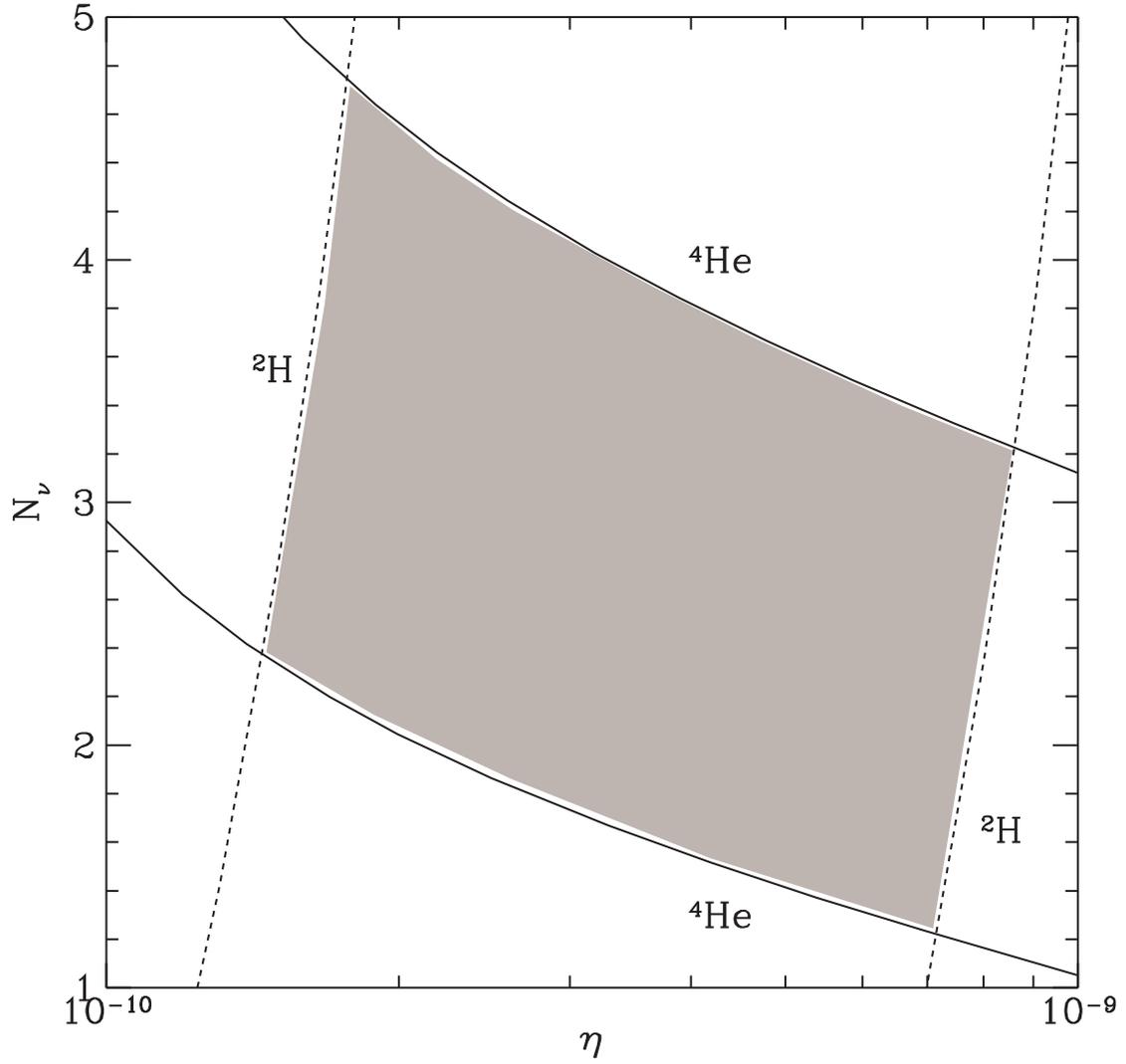}
\caption{The allowed region in the 
	$\eta-N_{\nu}$ plane for
	conservative estimates of the primordial abundances: 
	$d_4 \le 2.3$ (Rugers \& Hogan 1996), $d_5 \ge 1.5$ (Tytler,
	Fan, \& Burles 1996); 
	$Y \le 0.255$ (Sasselov \& Goldwirth 1995),
	$Y \ge 0.223$ (Olive 1995).} 
\end{figure}	
	 
\end{document}